\documentclass[conference]{IEEEtran}
\IEEEoverridecommandlockouts

\usepackage{amsmath,amssymb,amsfonts}
\usepackage{algorithmic}
\usepackage{graphicx}
\usepackage{textcomp}
\usepackage{svg}
\usepackage{url}
\usepackage{colortbl}
\usepackage{hhline}
\usepackage{relsize}
\usepackage[hidelinks,colorlinks=true,linkcolor=blue,citecolor=cyan]{hyperref}
\usepackage{pifont}

\usepackage{balance}
\usepackage{cite}

\usepackage{hhline}
\usepackage{arydshln}
\usepackage{textcomp}

\usepackage{xcolor}
\usepackage{ifthen}
\usepackage{tikz}
\usepackage{pgfplots,pgfplotstable}
\pgfplotsset{
        set layers={
            background,
            main,
        },
    }
\newlength{\figheight}
\usepackage[square,sort,comma,numbers]{natbib}
\usepackage[T1]{fontenc}

\usetikzlibrary{calc, patterns, spy}

\usepackage{listings}

\usepackage[binary-units=true]{siunitx} 
\usepackage[frozencache,cachedir=minted-cache]{minted}
\usepackage{listings}
\setminted{fontsize=\small,baselinestretch=1}
\usepackage{booktabs}
\setcitestyle{square}
\usepackage{multirow}
\usepackage{ragged2e}
\usepackage{pifont}
\usepackage{courier}
\usepackage{comment}

\usepackage{circledsteps}
\newcommand\myCircled[2][]{\ifmmode
\Circled[fill color=black,inner color=white,#1]{\mathsf{#2}}
\else
\Circled[fill color=black,inner color=white,#1]{\sffamily#2}
\fi
}

\usepackage{pifont}

\newcommand\encircle[1]{%
\tikz[baseline=(X.base),font=\sffamily] 
  \node (X) [draw, scale=0.7, shape=circle, inner sep=0, fill=black, text=white, minimum size=0em] {\strut #1};}

\usepackage[ruled, lined, linesnumbered, commentsnumbered, longend]{algorithm2e}

\include{acronyms}

\usepackage{svg}

\usepackage{amsmath}
\usepackage{nicematrix}
\definecolor{pairedOneLightBlue}{HTML}{a6cee3}
\definecolor{pairedTwoDarkBlue}{HTML}{1f78b4}
\definecolor{pairedThreeLightGreen}{HTML}{b2df8a}
\definecolor{pairedFourDarkGreen}{HTML}{33a02c}
\definecolor{pairedFiveLightRed}{HTML}{fb9a99}
\definecolor{pairedSixDarkRed}{HTML}{e31a1c}
\definecolor{butter1}{rgb}{0.988,0.914,0.310}
\definecolor{butter2}{rgb}{0.929,0.831,0.000}
\definecolor{butter3}{rgb}{0.769,0.627,0.000}
\definecolor{orange1}{rgb}{0.988,0.686,0.243}
\definecolor{orange2}{rgb}{0.961,0.475,0.000}
\definecolor{orange3}{rgb}{0.808,0.361,0.000}
\definecolor{chocolate1}{rgb}{0.914,0.725,0.431}
\definecolor{chocolate2}{rgb}{0.757,0.490,0.067}
\definecolor{chocolate3}{rgb}{0.561,0.349,0.008}
\definecolor{chameleon1}{rgb}{0.541,0.886,0.204}
\definecolor{chameleon2}{rgb}{0.451,0.824,0.086}
\definecolor{chameleon3}{rgb}{0.306,0.604,0.024}
\definecolor{skyblue1}{rgb}{0.447,0.624,0.812}
\definecolor{skyblue2}{rgb}{0.204,0.396,0.643}
\definecolor{skyblue3}{rgb}{0.125,0.290,0.529}
\definecolor{plum1}{rgb}{0.678,0.498,0.659}
\definecolor{plum2}{rgb}{0.459,0.314,0.482}
\definecolor{plum3}{rgb}{0.361,0.208,0.400}
\definecolor{scarletred1}{rgb}{0.937,0.161,0.161}
\definecolor{scarletred2}{rgb}{0.800,0.000,0.000}
\definecolor{scarletred3}{rgb}{0.643,0.000,0.000}
\definecolor{aluminium1}{rgb}{0.933,0.933,0.925}
\definecolor{aluminium2}{rgb}{0.827,0.843,0.812}
\definecolor{aluminium3}{rgb}{0.729,0.741,0.714}
\definecolor{aluminium4}{rgb}{0.533,0.541,0.522}
\definecolor{aluminium5}{rgb}{0.333,0.341,0.325}
\definecolor{aluminium6}{rgb}{0.180,0.204,0.212}

\definecolor{blind_safe_one_scheme_three_colors}{RGB}{102,194,165}
\definecolor{blind_safe_two_scheme_three_colors}{RGB}{252,141,98}
\definecolor{blind_safe_three_scheme_three_colors}{RGB}{141,160,203}

\definecolor{blind_safe_one_scheme_four_colors}{RGB}{166,206,227}
\definecolor{blind_safe_two_scheme_four_colors}{RGB}{31,120,180}
\definecolor{blind_safe_three_scheme_four_colors}{RGB}{178,223,138}
\definecolor{blind_safe_four_scheme_four_colors}{RGB}{51,160,44}

\definecolor{blind_safe_one_scheme_five_colors}{RGB}{240,249,232}
\definecolor{blind_safe_two_scheme_five_colors}{RGB}{186,228,188}
\definecolor{blind_safe_three_scheme_five_colors}{RGB}{123,204,196}
\definecolor{blind_safe_four_scheme_five_colors}{RGB}{67,162,202}
\definecolor{blind_safe_five_scheme_five_colors}{RGB}{8,104,172}

\definecolor{blind_safe_one_scheme_eight_colors}{RGB}{247,252,240}
\definecolor{blind_safe_two_scheme_eight_colors}{RGB}{224,243,219
}
\definecolor{blind_safe_three_scheme_eight_colors}{RGB}{204,235,197}
\definecolor{blind_safe_four_scheme_eight_colors}{RGB}{168,221,181}
\definecolor{blind_safe_five_scheme_eight_colors}{RGB}{123,204,196}
\definecolor{blind_safe_six_scheme_eight_colors}{RGB}{78,179,211}
\definecolor{blind_safe_seven_scheme_eight_colors}{RGB}{43,140,190}
\definecolor{blind_safe_eight_scheme_eight_colors}{RGB}{8,88,158}

\definecolor{blind_safe_one_scheme_seven_colors}{RGB}{118,42,131}
\definecolor{blind_safe_two_scheme_seven_colors}{RGB}{175,141,195}
\definecolor{blind_safe_three_scheme_seven_colors}{RGB}{231,212,232}
\definecolor{blind_safe_four_scheme_seven_colors}{RGB}{247,247,247}
\definecolor{blind_safe_five_scheme_seven_colors}{RGB}{217,240,211}
\definecolor{blind_safe_six_scheme_seven_colors}{RGB}{127,191,123}
\definecolor{blind_safe_seven_scheme_seven_colors}{RGB}{27,120,55}

\definecolor{yellow_one}{RGB}{255,255,212}
\definecolor{yellow_two}{RGB}{254,217,142}
\definecolor{yellow_three}{RGB}{254,153,41}
\definecolor{yellow_four}{RGB}{217,95,14}
\definecolor{yellow_five}{RGB}{153,52,4}

\definecolor{forestgreen}{RGB}{34, 139, 34}

\definecolor{hpca_one_of_ten_colors}{RGB}{158,1,66}
\definecolor{hpca_two_of_ten_colors}{RGB}{213,62,79}
\definecolor{hpca_three_of_ten_colors}{RGB}{244,109,67}
\definecolor{hpca_four_of_ten_colors}{RGB}{253,174,97}
\definecolor{hpca_five_of_ten_colors}{RGB}{254,224,139}
\definecolor{hpca_six_of_ten_colors}{RGB}{230,245,152}
\definecolor{hpca_seven_of_ten_colors}{RGB}{171,221,164}
\definecolor{hpca_eight_of_ten_colors}{RGB}{102,194,165}
\definecolor{hpca_nine_of_ten_colors}{RGB}{50,136,189}
\definecolor{hpca_ten_of_ten_colors}{RGB}{94,79,162}

\newcommand{\red}{}

\renewcommand{\red}{\textcolor{red}}

\begin{document}

\title{\huge {All-in-Memory Stochastic Computing using ReRAM
\vspace{-0.5em}
}}

\author{\IEEEauthorblockN{João Paulo C. de Lima\textsuperscript{1,2}, Mehran Shoushtari Moghadam\textsuperscript{3},\\ Sercan Aygun\textsuperscript{4}, Jeronimo Castrillon\textsuperscript{1,2,5}, M. Hassan Najafi\textsuperscript{3}, and Asif Ali Khan\textsuperscript{1}}
\IEEEauthorblockA{\textsuperscript{1}Chair for Compiler Construction, Dresden University of Technology, Dresden, Germany\\
 \textsuperscript{2}Center for Scalable Data Analytics and Artificial Intelligence (ScaDS.AI), Dresden, Germany \\ 
  \textsuperscript{3}Electrical, Computer, and Systems Engineering Department, Case Western Reserve University, Cleveland, OH, USA \\
  \textsuperscript{4}School of Computing and Informatics, University of Louisiana at Lafayette, Lafayette, LA, USA \\ 
  \textsuperscript{5}Barkhausen Institut, Dresden, Germany \\ 
Corresponding authors: \{joao.lima, asif\_ali.khan\}@tu-dresden.de}
\vspace{-1.5em}
}

\maketitle

\begin{abstract}
As the demand for efficient, low-power computing in embedded and edge devices grows, traditional computing methods are becoming less effective for handling complex tasks.
Stochastic computing (SC) offers a promising alternative by approximating complex arithmetic operations, such as addition and multiplication, using simple 
bitwise operations, like \texttt{majority} or \texttt{AND}, on random bit-streams.
While SC operations are inherently fault-tolerant, their accuracy largely depends on the length and quality of the stochastic bit-streams (SBS). These bit-streams are typically generated by CMOS-based stochastic bit-stream generators that 
consume over 80\% of the SC system's power and area.
Current SC solutions focus on optimizing the logic gates but often neglect the high cost of moving the bit-streams between memory and processor. This work leverages the physics of emerging ReRAM devices to implement the entire SC flow in place: \ding{182}~generating low-cost true random numbers and SBSs, \ding{183}~conducting SC operations, and \ding{184}~converting SBSs back to binary. Considering the low reliability of ReRAM cells, we demonstrate how SC's robustness to errors copes with ReRAM's variability. Our evaluation shows significant improvements in throughput (1.39$\times$, 2.16$\times$) and energy consumption (1.15$\times$, 2.8$\times$) over state-of-the-art (CMOS- and ReRAM-based) solutions, respectively, with an average image quality drop of 5\% across multiple SBS lengths and image processing tasks.
\end{abstract}

\section{Introduction}
\label{sec:intro}
The growing prevalence of embedded and edge devices has driven the demand for low-cost but efficient computing solutions. These devices, which often run complex applications like computer vision tasks in real-world environments, are constrained by computational resources and power budget, making traditional computing methods less effective. Stochastic computing (SC) and non-von Neumann paradigms have emerged as promising alternatives, offering 
trade-offs in computational density, energy efficiency, and error tolerance~\cite{frasser2022fully, daniels2020energy, sebastian2020memory, Alaghi_Survey_2018}.

In SC, data is represented by random bit-streams, where a value $x\!\in\! [0,1]$ is encoded by the probability ($P_x$) of a `1' appearing in the stream. For example, the bit-stream `10101' represents the value $\frac{3}{5}$, where $5$ is the bit-stream length ($N$). 
This unconventional representation enables complex computations like \emph{multiplication} and \emph{addition} to be approximated with 
simple logic operations such as \texttt{AND} and \texttt{majority}, respectively, reducing area and power consumption substantially without taking a high toll on computational accuracy. 
Additionally, since all bits in the bit-streams carry equal weights -- no \textit{most}- or \textit{least}-significant bits -- SC is naturally tolerant to noise, including bit flips and inaccuracies in the input data and computations. This makes SC particularly advantageous for a range of applications, including image processing~\cite{li2013computation}, signal processing~\cite{chang2013architectures}, and neural networks~\cite{ren2017sc}.

Stochastic bit-streams (SBSs) are conventionally generated using a CMOS-based structure, called stochastic bit-stream generator, built from a \emph{pseudo random} (or more recently, \emph{\textit{quasi-random})}~\cite{8327916} \emph{number generator}, and a \emph{binary comparator}. 
The accuracy and cost efficiency of SC systems highly depend on this bit-stream generation unit. 
Presently, CMOS-based bit-stream generation 
consumes up to 80\% of the system's total hardware cost and energy consumption~\cite{Najafi_TVLSI_2019,Alaghi_Survey_2018}. 
Additionally, SC implementations on classic von Neumann systems require extensive movement of bit-streams \textit{from}\slash \textit{to} memory, which is often overlooked in evaluations but can easily offset the benefits of the simpler SC operations. 
This has motivated significant research into non-von Neumann computing for SC, using different memory technologies~\cite{SC_InMemory_Native2014, SC_InMemory_2014, SC_Memristor_2017, SCRIMP_DATE20,Unary_DATE21,Unary_IMC_2021,IMC_SC_DNN_2021,RiahiAlam_DT2021_}.

\begin{figure}[t]
    \centering
    \includegraphics[width=0.9\columnwidth]{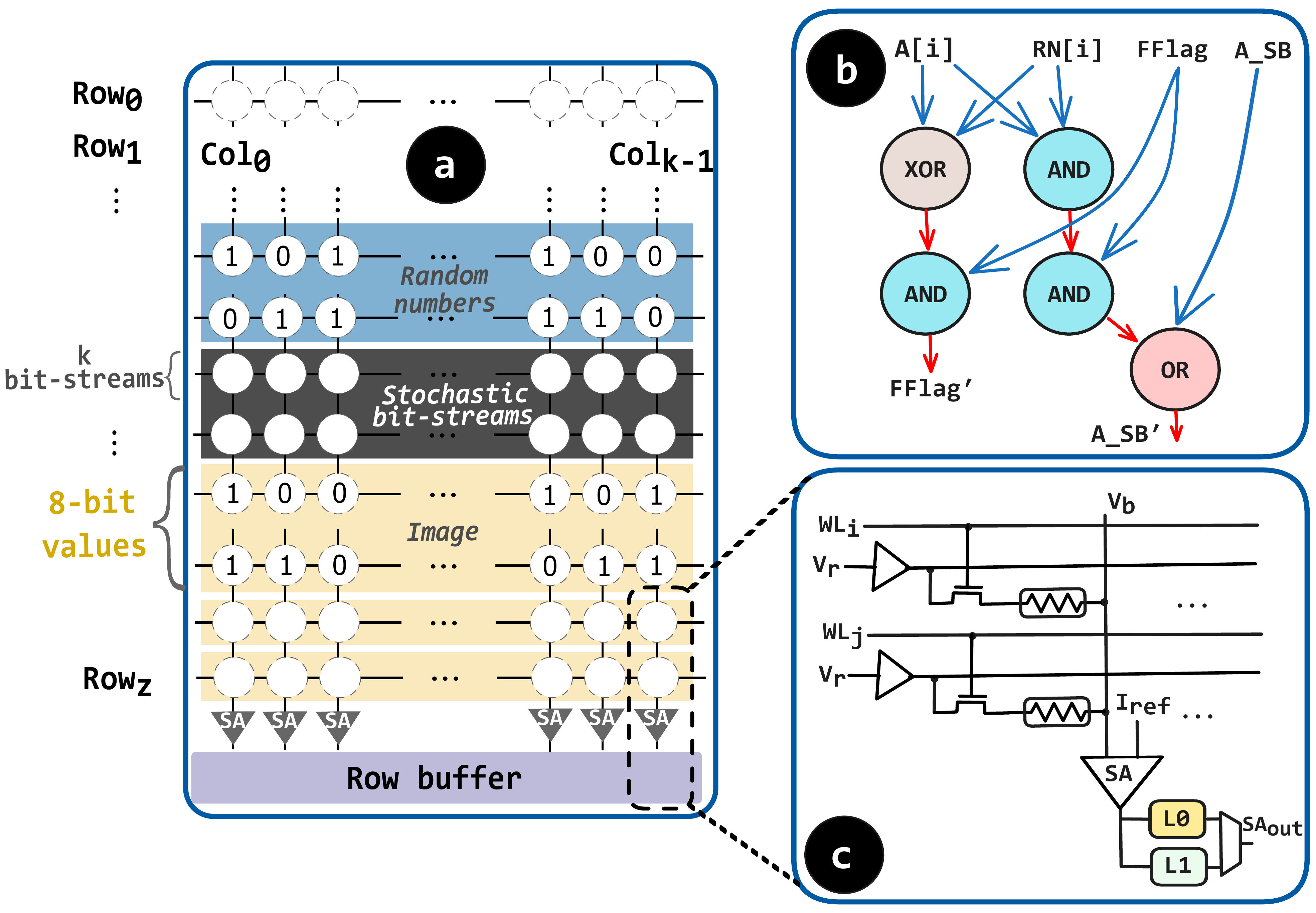}
    \vspace{-0.5em}
    \caption{A high-level overview of our proposed in-memory SC solution: 
    (a) ReRAM array, (b) Greater-than operation using basic logic gates, (c) Write latches in the peripheral circuitry.
    }
    \label{fig:overview}
    \vspace{-1.2em}
\end{figure}

Most compute-in-memory (CIM) designs for SC utilize ReRAM, a nonvolatile memory technology that stores data in the resistance state of the devices~\cite{reram}. ReRAM offers DRAM-comparable read latency, higher density, and significantly reduced read energy consumption but incurs expensive write operations, limited write endurance, and suffers from non-linearities~\cite{reram-challenges}.
SC in ReRAM benefits from ReRAM’s high density for efficient storage and in-place processing of long bit-streams, while SC’s inherent robustness helps mitigate the effects of ReRAM’s non-linearities. In existing 
ReRAM-based CIM designs for SC, conventional CMOS-based random number generators (RNGs) are used for bit-stream generation, while ReRAM arrays handled in-place logic operations~\cite{SC_InMemory_Native2014,SC_InMemory_2014,SC_Memristor_2017}. This increases parallelism and reduces data movement overhead, though the high cost of random bit-stream and SBSs generation remains a bottleneck. 
Techniques like leveraging ReRAM’s inherent write noise~\cite{SCRIMP_DATE20} and employing DRAM-based lookup tables~\cite{sc-in-DRAM} have been explored to improve the performance of random bit-stream generation. However, SBSs generation continues to face challenges, including the high energy cost of ReRAM write operations and scalability issues with DRAM-based methods.

To address these challenges, we propose a 
CIM accelerator that implements \emph{all steps} of SC using ReRAM. We decouple RNG from SBS generation, allowing compatibility with any RNG type, including general-purpose true RNGs (TRNGs) based on ReRAM~\cite{Woo2019}. This approach ensures (1)~\textit{accurate SBS generation} with target probabilities and (2)~\textit{correlation control}, despite ReRAM cell noise.
We perform in-memory logic and comparison operations to convert true random numbers into SBS (\ding{182}), conduct SC operations (\ding{183}), and convert the resulting values back into a binary representation (\ding{184}). 
Concretely, this work makes the following novel contributions:

\begin{itemize} 
\item
We propose a ReRAM-based accelerator for SC that implements all steps in place, including those often overlooked by current SC designs. 

\item We develop a novel in-memory method for converting true random binary sequences (50\% \textit{ones}) into SBSs with desired probabilities. To the best of our knowledge, this is the first such method reported in the literature. 

\item 
Our SBS generation approach is RNG-agnostic, leveraging in-memory comparison to produce SBS even under substantial CIM failures caused by ReRAM variability.

\item
For SC operations typically implemented with \texttt{MUX}s, we propose novel alternatives that are CIM-friendly and achieve comparable accuracy. 

\end{itemize}
 
Compared to the state-of-the-art CMOS-based solutions, the proposed design, while requiring minimal changes to the memory periphery, on average, reduces energy consumption by $1.15\times$ and improves throughput by $1.39\times$ across multiple image processing applications. Our design is also more robust than traditional arithmetic for CIM, with only a 5\% average quality drop in the presence of faults compared to a 47\% drop. It eliminates the need for protection schemes on unreliable ReRAM devices and provides better correlation control than previous in-memory SC designs. 
\section{Background and Related Work}
\label{sec:bg}

\subsection{ReRAM-based Computing}
\label{subsec:reram-bg}
Resistive RAM (ReRAM) is a type of nonvolatile memory where each cell, typically a metal oxide, and being referred to as a memristor, is programmed to different resistance states using an electric voltage~\cite{dittmann2021nanoionic}. Data is represented through resistance levels, such as a high resistance state (HRS) for `0' and a low resistance state (LRS) for `1'. Organized in a conventional memories like 2D grid of rows (wordlines, WL) and columns (bitlines, BL), ReRAM promises DRAM-comparable read performance but has costly write operations that impact both energy consumption and the write endurance~\cite{reram-challenges}. 
For CIM using ReRAM, the 1T1R (one transistor, one resistor) crossbars are extensively used in machine learning and other domains to perform analog matrix-vector multiplication in constant time~\cite{sebastian2020memory}. 
Similarly, stateful and non-stateful logic techniques, such as MAGIC~\cite{kvatinsky2014magic} and scouting logic (SL)~\cite{xie2017scouting}, respectively, have been demonstrated for implementing logic operations using ReRAM.
ReRAM cells have inherent stochasticity and noise; which have also been investigated to generate true random numbers~\cite{Woo2019, Schnieders_2024}. 

\subsection{Stochastic Computing (SC)}
\label{subsec:sc-bg}
SC is an alternate computing approach offering simple execution of complex arithmetic operations and high tolerance to soft errors. 
Unlike traditional binary radix, SC operates on random bit-streams of `0's and `1's, with no bit-significance. 
SC systems include three primary components: \ding{182} Bit-stream generator that converts data from traditional binary to stochastic bit-stream, \ding{183} computation logic that performs bit-wise operations on the bit-streams, and \ding{184} bit-stream to binary converter to convert data back to binary format.

\noindent\textbf{\textit{Bit-stream Generation:}} The accuracy of SC operations highly depends on the quality of bit-streams.
To convert a binary number $X$ to an SBS of size $N$, an 
RNG is used to generate $N$ random numbers. The SBS is generated by comparing each of these $N$ random numbers with $X$. A `1' is produced if the random number is less than  $X$, and a `0' is produced otherwise. Conventionally, SC systems employ CMOS-based pseudo-RNGs (PRNGs) such as linear-feedback shift registers (LFSRs) to generate the needed random numbers~\cite{Weikang_2011}. 
However, this can lead to suboptimal performance as 
very long SBSs are needed to achieve acceptable accuracy. Recent works leverage quasi-RNGs (QRNGs) for better accuracy~\cite{p2lsg} but at the cost of a higher area and power~\cite{8327916, Najafi_TVLSI_2019}. The high cost of CMOS-based SBS generation offsets the gains made with simple computation circuits. 

\noindent\textbf{\textit{SC Operations:}} Basic arithmetic operations -- multiplication, addition, subtraction, and division -- are implemented in SC using minimal components: an \texttt{AND} gate, a multiplexer (\texttt{MUX}) unit, an \texttt{XOR} gate, and a \texttt{MUX}+\texttt{D}-flip-flop, respectively (Fig.~\ref{Fig:TRNG_SC})~\cite{Alaghi_Survey_2018,CORDIV}. For $N$-bit-long SBSs, the logic operations are often performed serially, producing the output SBS in $N$ clock cycles. Parallel execution of the operations is also feasible by trading off time with space. 
This approach is particularly attractive for SC with CIM as it enables 
fast and independent execution of all bit-wise operations. 
For correct functionality of the aforementioned operations, the input bit-streams must 
provide the desired correlation level,
i.e., \textit{uncorrelated} for the multiplication and addition, and \textit{correlated} for the subtraction and division operations. The independence (i.e., uncorrelation) requirement 
is typically satisfied by using different RNGs while the desired amount of correlation 
is guaranteed by using shared RNGs when generating SBSs. 

\subsection{State-of-the-art In-memory SC Solutions}
\label{subsec:related-work}
Existing CIM-SC designs are mostly hybrid, i.e., either memristive arrays are used to generate random numbers and CMOS logic to perform computations, or vice versa. For instance, Knag et al.~\cite{SC_InMemory_Native2014} proposed generating SBSs using memristors and off-memory computations using CMOS stochastic circuits. A similar design is proposed in~\cite{IMC_SC_DNN_2021} that exploits the switching stochasticity of probabilistic Conductive Bridging RAM (CBRAM) devices to generate SBSs in memory efficiently. The generated bit-streams are then used to optimize deep learning parameters using a hybrid CMOS-memristor stochastic processor.
ReRAM-based SBS generation has also been proposed for SC. However, these solutions primarily use the probabilistic switching, i.e., the write operation in ReRAM~\cite{Physic_RRAM_SC_SNG_2019}. Riahi Alam et al.~\cite{RiahiAlam_DT2021_} developed an accurate method for in-memory SC multiplication by performing a deterministic binary-to-SBSs conversion. Sun~et~al.~\cite{Unary_IMC_2021} employed unary coding, using multi-level memristor cells, for weight representation in a ReRAM-based neural network accelerator. The most relevant work to our design is SCRIMP~\cite{SCRIMP_DATE20}, which also proposes SBS generation and computation using ReRAM. However, similar to other prior works, it generates SBSs using the stochasticity in the write operation, which is not only extremely slow but also affects write endurance.
Existing methods can generate SBSs with target probabilities but \textit{lack correlation control}, which limits their applicability for SC operations that require correlated inputs.
We propose a novel ReRAM-based solution to convert true random binary sequences (50\% \textit{ones}) into SBSs with desired probabilities using bulk-bitwise operations.

\section{Proposed In-ReRAM Stochastic Computing}
\label{sec:methods}
We exploit the physical properties of ReRAM arrays to implement all stages of the SC flow (see Sec.~\ref{subsec:sc-bg}). In practice, we use multiple arrays to parallelize and pipeline the different stages. However, for simplicity, we show a single array in Fig.~\ref{fig:overview}\encircle{a}, consisting of dedicated rows to store input binary data (\textcolor{black}{yellow}), random numbers (blue), and in-memory generated stochastic bit-streams (grey). In the following, we explain the in-memory implementation of these different operations and the data flow in the different stages of the SC flow.

\subsection{Stochastic Number Generation (SNG)}
\label{subsec:stoch-num-gen}


The switching stochasticity of ReRAM devices has been exploited to generate true random numbers (see Sec.~\ref{subsec:reram-bg}). We build upon this prior work and consider TRNG as a single-step operation that stores random sequences directly in ReRAM arrays. To generate SBSs from these random sequences (referred to as in-memory SNG or IMSNG), we compare them with $n$-bit input binary operands using in-memory bitwise operations (see Sec.~\ref{subsec:sc-bg}).
Concretely, for comparing two binary numbers $A$ and $RN$ in memory, starting from the most significant bit (MSB) to the least significant bit (LSB), we perform bitwise comparison and stop at the first non-equal bit position, i.e., where $A_i \oplus RN_i$ is 1. This is achieved by implementing the \texttt{greater-than} operation, i.e., $A_i\!\!>\!\!RN_i$, using in-ReRAM bitwise \texttt{XOR} and \texttt{AND} operation, together with a flag bit (\texttt{FFlag}), as illustrated in the Boolean network in Fig.~\ref{fig:overview}\encircle{b}. The result of this comparison is a row $A_{SB}$ representing the SBS of $A$. 
This network is converted into data structures like \texttt{XOR}-\texttt{AND}-Inverter graph (XAG) for manipulation and optimization using logic synthesis tools~\cite{soeken2018epfl}.

\begin{figure}[t]
		\begin{center}
\includegraphics[width=\linewidth,   keepaspectratio,
			]{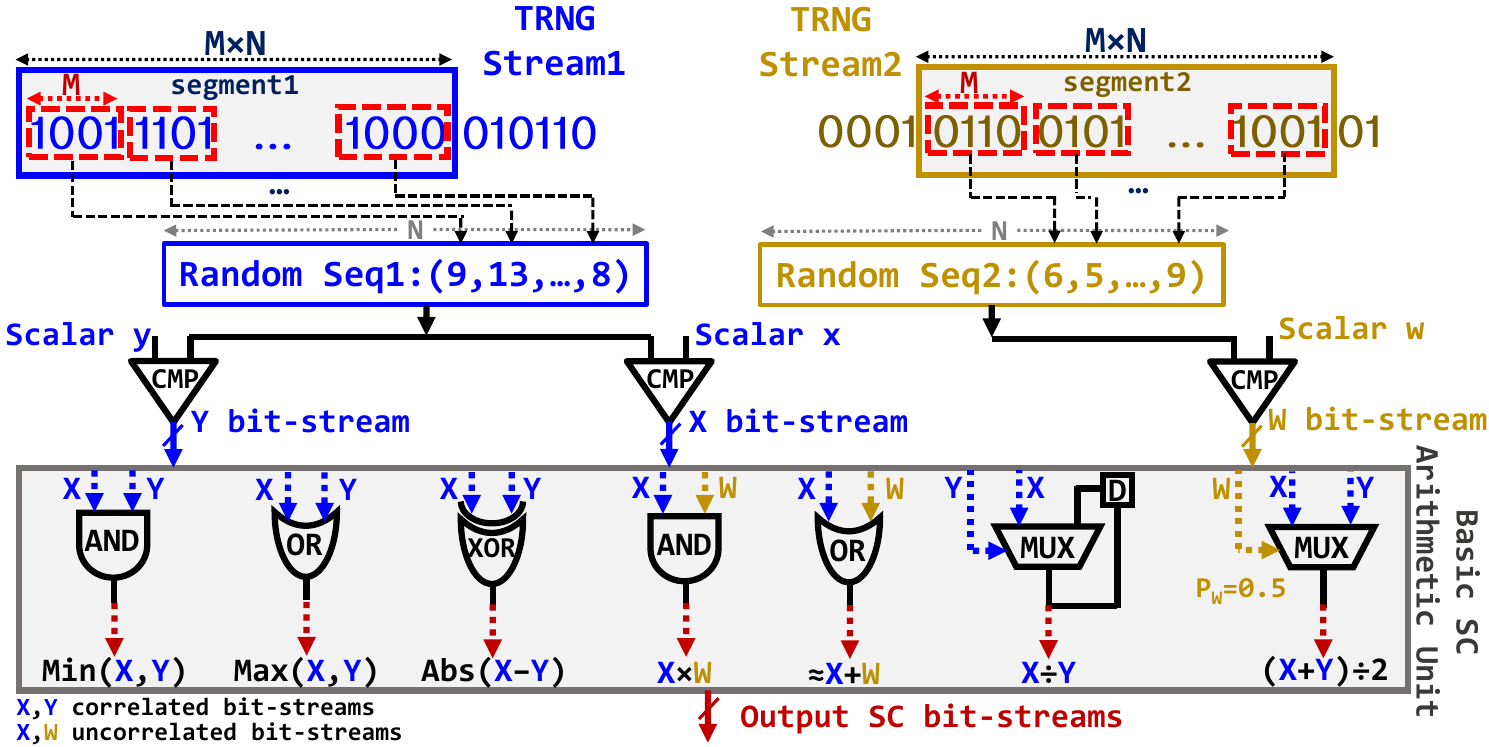}
   \vspace{-1.8em}
\caption {In-ReRAM SBS generation and SC arithmetic operations. CORDIV~\cite{CORDIV} is considered for SC division with $X\!\leqslant\!Y$. Addition with \texttt{OR}; the inputs are in the $[0,0.5]$ interval to not exceed $1.0$ in the output.}
\label{Fig:TRNG_SC}
		\end{center}
  \vspace{-1.em}
\end{figure}

\begin{table}[t]
\centering
\arrayrulecolor{black}
\caption{MSE(\%) comparison of generating SBSs utilizing different RNG sources (M: Block size, N: bit-stream length).}
 \vspace{-0.8em}
\setlength{\tabcolsep}{4.5pt}
 \renewcommand{\arraystretch}{1.1}
 \relscale{0.86}
\begin{tabular}{!{\color{black}\vrule}c|c!{\color{black}\vrule}ccccc!{\color{black}\vrule}} 
\hline
\multicolumn{2}{|c|}{\textbf{ RNG Source }} & \multicolumn{5}{c!{\color{black}\vrule}}{\textbf{ Bit-stream length (N) }} \\ 
\hline
\multirow{6}{*}{\begin{tabular}[c]{@{}c@{}} \textbf{IMSNG}\\ \cite{Woo2019} \end{tabular}} & \begin{tabular}[c]{@{}c@{}} \textbf{ Segment size} \textbf{(M)} \end{tabular} & \textbf{ 32 } & \textbf{ 64 } & \textbf{ 128 } & \textbf{ 256 } & \textbf{ 512 } \\ 
\arrayrulecolor{black}\cline{2-6}\arrayrulecolor{black}\cline{7-7}
 & \textbf{ 5 } & 0.567 & 0.321 & 0.189 & 0.134 & 0.103 \\
 & \textbf{ 6 } & 0.562 & 0.302 & 0.177 & 0.114 & 0.084 \\
 & \textbf{ 7 } & 0.534 & 0.279 & 0.157 & 0.095 & 0.064 \\
 & \textbf{ 8 } & 0.557 & 0.300 & 0.177 & 0.107 & 0.074 \\
 & \textbf{ 9 } & 0.520 & 0.282 & 0.159 & 0.090 & 0.060 \\ 
\cline{1-2}\arrayrulecolor{black}\cline{3-6}\arrayrulecolor{black}\cline{7-7}
\multicolumn{2}{|c|}{\textbf{ Software - MATLAB }} & 0.529 & 0.264 & 0.131 & 0.065 & 0.032 \\
\multicolumn{2}{|c|}{\textbf{ PRNG (8-Bit LFSR) }} & 1.069 & 0.554 & 0.288 & 0.137 & 0.071 \\
\multicolumn{2}{|c|}{\textbf{ QRNG (8-Bit Sobol) }} & 0.033 & 0.008 & 0.002 & 5.05\texttimes 10\textsuperscript{-4} & 1.25\texttimes 10\textsuperscript{-4} \\ 
\cline{1-2}\arrayrulecolor{black}\cline{3-6}\arrayrulecolor{black}\cline{7-7}
\end{tabular}
\label{tab:SNG-Accuracy}
\vspace{-0.5em}
\justify{\scriptsize{
For PRNG, a Maximal length LFSR with polynomial $x^8+x^5+x^3+1$ is used.
}}
\vspace{-1em}
\end{table}

Considering the SL approach (see Sec.~\ref{subsec:reram-bg}), implementing this network requires $5n$ operations, as each logic gate requires one sensing step (cf.~\cite{xie2017scouting}). This also requires writing the intermediate signals (e.g., that of \texttt{XOR}) back to the ReRAM array at least 4 times (red arrows in Fig.~\ref{fig:overview}\encircle{b}). These intermediate writes can be avoided by forwarding the output of one operation directly as input voltage to the bitline for the next operation. This requires the periphery to incorporate a simple feedback mechanism that converts the latched signal to adjust the bitline voltage (\si{\volt}$_b$) (see Fig.~\ref{fig:overview}\encircle{c}) to mimic the voltage drop across the ReRAM that would have been written. 
This approach (referred to as \emph{IMSNG-naive}) reduces the number of ReRAM writes to $2n$. 

\begin{table*}
\centering
\caption{MSE (\%) comparison of SC arithmetic operations utilizing different RNGs with $M=8$.}
\vspace{-1em}
\setlength{\tabcolsep}{3.2pt}
\relscale{0.825}
\label{MAE_op}
\begin{tabular}{|c|ccccc|ccccc|ccccc|ccccc|} 
\hline
\multirow{2}{*}{\begin{tabular}[c]{@{}c@{}}\textbf{SC}\\ \textbf{Operations}\end{tabular}} & \multicolumn{5}{c|}{\textbf{IMSNG}~\cite{Woo2019}} & \multicolumn{5}{c|}{\textbf{ Software - MATLAB }} & \multicolumn{5}{c|}{\textbf{ PRNG (LFSR) }} & \multicolumn{5}{c|}{\textbf{ QRNG (Sobol) }} \\ 
\cline{2-21}
 & \textbf{ N:32 } & \textbf{ 64 } & \textbf{ 128 } & \textbf{ 256 } & \textbf{ 512 } & \textbf{ N:32 } & \textbf{ 64 } & \textbf{ 128 } & \textbf{ 256 } & \textbf{ 512 } & \textbf{ N:32 } & \textbf{ 64 } & \textbf{ 128 } & \textbf{ 256 } & \textbf{ 512 } & \textbf{ N:32 } & \textbf{ 64 } & \textbf{ 128 } & \textbf{ 256 } & \textbf{ 512 } \\ 
\hline
\textbf{ Multiplication } & 0.473 & 0.255 & 0.147 & 0.091 & 0.061 & 0.444 & 0.219 & 0.108 & 0.054 & 0.027 & 0.851 & 0.476 & 0.221 & 0.093 & 0.060 & 0.058 & 0.017 & 0.005 & 0.001 & 2.9$\times$10\textsuperscript{-4} \\
\textbf{Scaled Addition } & 0.690 & 0.356 & 0.193 & 0.109 & 0.062 & 0.648 & 0.328 & 0.159 & 0.082 & 0.041 & 1.117 & 0.607 & 0.289 & 0.157 & 0.065 & 0.102 & 0.013 & 0.003 & 0.002 & 2.1$\times$10\textsuperscript{-4} \\
\textbf{Approx. Addition } & 1.548 & 1.186 & 1.024 & 0.927 & 0.886 & 1.379 & 1.055 & 0.897 & 0.789 & 0.751 & 2.654 & 1.702 & 1.180 & 0.914 & 0.842 & 0.463 & 0.586 & 0.670 & 0.662 & 0.689 \\
\textbf{Abs. Subtraction } & 0.641 & 0.354 & 0.136 & 0.144 & 0.107 & 0.514 & 0.263 & 0.129 & 0.064 & 0.034 & 0.559 & 0.281 & 0.136 & 0.058 & 0.026 & 0.016 & 0.004 & 0.001 & 2.5$\times$10\textsuperscript{-4} & 6.5$\times$10\textsuperscript{-5} \\
\textbf{ Division } & 1.614 & 0.895 & 0.518 & 0.295 & 0.187 & 1.454 & 0.789 & 0.392 & 0.196 & 0.106 & 2.760 & 2.140 & 1.688 & 1.630 & 1.477 & 0.251 & 0.164 & 0.129 & 0.126 & 0.128 \\
\textbf{ Minimum } & 0.572 & 0.307 & 0.177 & 0.106 & 0.064 & 0.514 & 0.265 & 0.130 & 0.066 & 0.032 & 1.493 & 0.811 & 0.394 & 0.199 & 0.085 & 0.033 & 0.008 & 0.002 & 5.1$\times$10\textsuperscript{-4} & 1.3$\times$10\textsuperscript{-4} \\
\textbf{ Maximum } & 0.572 & 0.302 & 0.186 & 0.117 & 0.077 & 0.543 & 0.259 & 0.132 & 0.064 & 0.033 & 0.481 & 0.263 & 0.123 & 0.073 & 0.027 & 0.032 & 0.008 & 0.002 & 5.0$\times$10\textsuperscript{-4} & 1.3$\times$10\textsuperscript{-4} \\
\hline
\end{tabular}
\vspace{-0.5em}
\end{table*}

\noindent \textbf{IMSNG-opt:} As an alternative approach, we are exploiting the latches in the ReRAM peripheral circuitry, shown in Fig.~\ref{fig:overview}\encircle{c} (L0 and L1), to minimize the write overhead. ReRAM and other nonvolatile memories, typically employ double latches and a write driver to conduct differential writes~\cite{chevallier20100}. 
For each write driver, a latch stores the data to be written to the cells and a second latch stores whether the cell should be modified (in case the new data is different than the already stored data). 
We leverage this mechanism to directly implement the \texttt{AND} operations involving \texttt{FFlag} as a predicated sensing, hence eliminating the need to write the intermediate result to the memory cells. This approach, which uses existing latches and drivers to eliminate write operations deriving from intermediate results, is referred to as \emph{IMSNG-opt}.

Table~\ref{tab:SNG-Accuracy} compares the mean squared error (MSE) of our IMSNG 
to a software-based (SW), PRNG-based, and QRNG-based SBS generator. SW uses MATLAB's RNG (\texttt{rand}) for bit-stream generation. PRNG and QRNG use 8-bit LFSR and Sobol sequence generators. 
The comparison in IMSNG is between the target input and $N$ random binary sequences (generated in-memory) of $M$ bits, where $M\!\!=\!\!5, 6, ..., 9$ (see Fig.~\ref{Fig:TRNG_SC}). The data is based on $1\!,\!000,\!000$ samples extracted from a uniform distribution. 
The results highlight that IMSNG, despite its random fluctuations and true randomness, provides comparable accuracy to other methods. Notably, for bit-stream lengths of 32, 64, and greater than 128, MSEs of approximately 0.5\%, 0.3\%, and 0.1\% are measured, respectively. 


\subsection{Stochastic Circuits using Scouting Logic (SL)}
\label{subsec:sc-ops}
SL implements boolean logic using ReRAM read operations with a modified sense amplifier (SA)~\cite{pinatubo,xie2017scouting}. 
During a logic operation, two or more rows are simultaneously activated and the resulting current through the cells in each bitline is compared with a reference current $I_{ref}$ by the SA, whose output is the desired result of the Boolean operation (see Fig.~\ref{fig:overview}\encircle{c}). All basic logic operations such as (\texttt{N})\texttt{AND}, (\texttt{N})\texttt{OR}, \texttt{X}(\texttt{N})\texttt{OR}, and \texttt{NOT}, are realized in a single cycle~\cite{yu2019enhanced}. 


Bulk bitwise in-memory operations are performed on large vectors. When operating on traditional binary-radix numbers, we can only exploit the single instruction, multiple data (SIMD) parallelism, since these algorithms are sequential by nature due to carry propagation. 
In contrast, SC handles basic arithmetic operations (addition, subtraction, multiplication, and division) using simple, low-cost logic units such as \texttt{AND}, \texttt{NOT}, \texttt{XOR}, \texttt{MUX}s, and flip-flops. Each bit is computed independently, allowing for in-memory SC to exploit bulk-bitwise logic and massive word-level parallelism, 
significantly reducing latency for basic arithmetic operations.
In the following, we explain how these primary arithmetic operations are implemented using bulk bitwise logic schemes.

\textbf{Multiplication} 
is implemented by performing bitwise \texttt{AND} on two independent bit-streams, representing probabilities $p$ and $q$. 
The probability of observing a `1' in the output stream equals $p \land q$, which aligns well with the principles of SL with the time complexity of $\mathcal{O}(1)$. 
This contrasts with conventional bulk-bitwise implementations of binary radix 
multiplication, which exhibit a time complexity of $\mathcal{O}(n^2)$, where $n$ represents the number of bits. This complexity arises from the iterative nature of traditional methods, which rely on bit shifts and additions to compute the product.

\textbf{Scaled addition} is 
implemented in SC using a 2-to-1 \texttt{MUX}.
In SL, a 2-to-1 \texttt{MUX} can be approximated by a CIM-friendly 3-input majority gate (\texttt{MAJ})~\cite{chen2015equivalence} that can be computed in a single cycle. 
For in-place \texttt{MAJ}, a reference current corresponding to the majority of the inputs is required. For instance, to conduct a 3-input \texttt{MAJ} gate operation, we use the same reference current used for the 2-input \texttt{AND} gate, as this detects when at least two out of three inputs are high.
The time complexity of our \texttt{MAJ}-based addition is $\mathcal{O}(1)$, which is a significant improvement over both traditional ripple-carry additions in the binary domain and \texttt{MUX}-based addition in existing SC that has a time complexity of $\mathcal{O}(N)$. 

\textbf{Division} in prior SC works 
is implemented using  CMOS-based \texttt{flip-flop}s and \texttt{MUX}s, and \textit{correlated} bit-streams to approximate $y = \frac{x_1}{x_2}$.
In SL, the \texttt{JK flip-flop}'s truth table can be implemented using the existing latch-based circuitry (Fig.~\ref{fig:overview}\encircle{c}). The intermediate values from the \texttt{flip-flop} are stored in the existing latch (write driver) and forwarded to the bitline as voltage inputs, eliminating the intermediate write operations and improving energy efficiency and write endurance. 
This method has a time complexity of $\mathcal{O}(N)$, while existing CIM division methods on integer data~\cite{leitersdorf2023aritpim} require $\mathcal{O}(n^2)$ write cycles.

For \textbf{other operations}, 
such as approximate addition, absolute subtraction, minimum, and maximum, we employ bulk-bitwise operations like \texttt{OR}, \texttt{XOR}, \texttt{AND}, and \texttt{OR}, respectively. In stochastic logic, the reference current for the \texttt{OR} operation is set to detect when at least one of the operands is high.

Table~\ref{MAE_op} compares the accuracy of these stochastic operations across different SNG sources. Compared to PRNG, QRNG, and SW-based SNG methods, our IMSNG approach achieves comparable accuracy, and even in some cases (e.g., compared to PRNG-based division) a lower MSE. Still, the important advantage of our method is that it is executed completely in memory, eliminating the overheads of transferring SBSs between memory and processing circuits.  

\subsection{Stochastic to Binary Conversion}
\label{subsec:sc2bin}
As a last step in the SC flow, the output of the SC operation needs to be converted back to binary. 
Existing methods use CMOS counters for stochastic-to-binary (S-to-B) conversion, sequentially counting the `1's in the output bit-stream. In contrast, our approach achieves the count in a single step by using bitline current accumulation. The output bit-stream is applied as input voltages ($V_r$) to a designated reference column in which all cells have been pre-programmed to low resistance states (see Fig.~\ref{fig:overview}\encircle{c}). The total current through the bitline, representing the population count of the bit-stream, is then measured and digitized using analog-to-digital converters (ADCs).



\section{Evaluation Results and Analysis}
\label{sec:eval}
The custom SA and the proposed hardware modifications, including the feedback mechanism and latch-based optimizations, were validated with SPICE simulations. Energy consumption and latency metrics of the in-memory design were extracted from~\cite{xie2017scouting} and integrated into NVMain~\cite{poremba2015nvmain}. For S-to-B conversion, we consider a single 8-bit ADC from~\cite{shafiee2016isaac} per mat. SL output is prone to failures when deciding on the bitwise operation output due to the intrinsic variability of ReRAM, as described in~\cite{farzaneh2024sherlock}.
We conduct simulations with the VCM-based ReRAM model~\cite{wiefels2020hrs} to determine the distribution of LRS and HRS that leads to the probability of obtaining incorrect outputs in CIM operation.
For NVMain simulation, we generate traces for the SBS generation, the SC circuits in Table~\ref{MAE_op}, and image processing applications. 
The derived failure rates are used to simulate fault injections and we report the average results of 1,000 runs.
For the CMOS-based SC circuits, we synthesized the designs using the Synopsys Design Compiler with the 45$nm$ gate library.

\subsection{Applications} 
\label{subsec:applications}
\begin{figure}[t]
\label{im_comp}
  \centering
  \includegraphics[width=\linewidth]{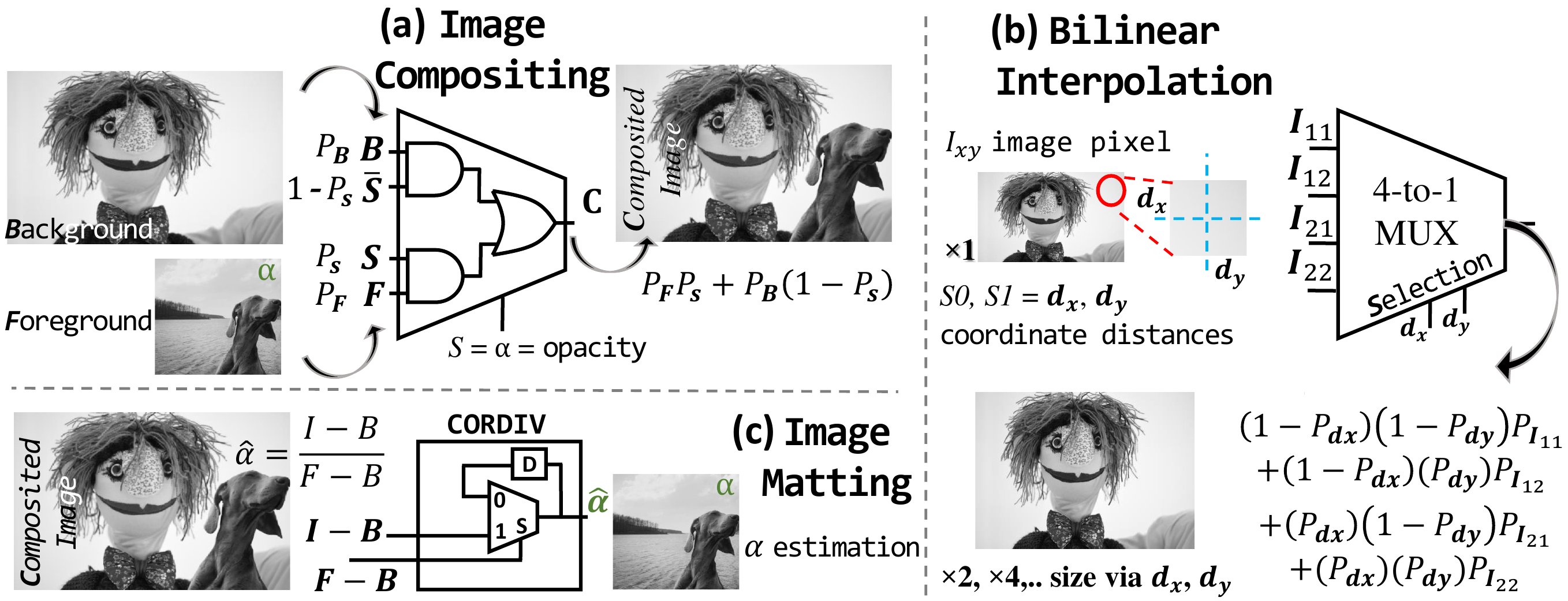}
  \vspace{-1.8em}
  \caption{SC image processing applications: (a) \textit{Image Compositing}, merging background and foreground images with $\alpha$ channel. (b) \textit{Bilinear Interpolation}, up-scaling input images. (c) \textit{Image Matting}, parsing $\alpha$ channel for the foreground object to separate the background.}
  \label{fig:ImageProc}
  \vspace{-0.5em}
\end{figure}


To evaluate our CIM-SC design, we use image processing applications, including (a) image compositing, (b) bilinear interpolation, and (c) image matting. 

\noindent\textbf{Image Compositing.} In traditional computer vision, two images — background ($B$) and foreground ($F$) — are merged using a linear formula 
that incorporates pixel information from both images, along with an additional channel, $\alpha$, which represents the opacity of the foreground. This process allows 
merging two scenes, typically when one of them has a green background. 
The $\alpha$-channel defines the object of interest, its refined edges, and the remaining background area in the final composite image. The compositing formula used, is $C = F \times \alpha + B \times (1 - \alpha)$, which corresponds to the \texttt{MUX} in the SC domain. If the bit-streams of $B$, $F$, and $\alpha$ are provided to the first, second, and selection inputs of the \texttt{MUX}, respectively, the output becomes the composite image, $C$. This process is illustrated in Fig.~\ref{fig:ImageProc}(a).

\textbf{Bilinear Interpolation.} Image up-scaling can be achieved through bilinear interpolation involving two linear interpolations along the \texttt{x}- and \texttt{y}-axes. For any 4-pixel neighboring group, $I_{11}$, $I_{12}$, $I_{21}$, and $I_{22}$, the input image $I$ can be used to create a larger image by estimating new pixel values between them. The intensity of the new pixel, $I(\mathrm{x},\mathrm{y})$, is calculated using the intensities of the four neighboring pixels and their relative distance ($dx, dy$) from the new point. The conventional formula is $I(\mathrm{x},\mathrm{y}) = (1-dx)(1-dy)I_{11} + (1-dx)(dy)I_{12} + (dx)(1-dy)I_{21} + (dx)(dy)I_{22}$. This corresponds to a 4-to-1 \texttt{MUX} in the SC domain. Here, the bit-streams of the four neighboring pixels serve as inputs to the \texttt{MUX}, while the $dx$ and $dy$ bit-streams are fed into the selection ports. The output bit-stream provides the intensity estimate for the new pixel in the up-scaled image. This application is shown in Fig.~\ref{fig:ImageProc}(b).

\textbf{Image Matting.} Image compositing can be reversed to separate the background and foreground images by estimating the $\alpha$-channel. By rearranging the composite image formula and solving for $\alpha$, the estimated alpha $\hat{\alpha} = \frac{I-B}{F-B}$ reveals the foreground object and helps refine the edges for a more natural composite image. This estimation is often repeated multiple times with different $B$ and $F$ information, especially when foreground details are missing. Given that this formula involves division operation~\cite{10415201}, this work employs the CORDIV design~\cite{CORDIV}, as shown in Fig.~\ref{fig:ImageProc}(c).

\subsection{Performance and Energy Comparison}
Table~\ref{HW_costs} compares the hardware costs of CMOS-based (\ding{59}) and ReRAM-based (\ding{70}) SC arithmetic designs. The analysis focuses on the breakdown of SC logic, excluding memory movement costs for CMOS-based designs that further increase their total latency and energy consumption. CMOS implementations employ LFSR and Sobol generators as SNG. Regarding latency, CMOS-based designs are significantly slower than ReRAM-based designs due to the sequential processing of bit-streams. Our ReRAM-based design reduces latency by 38\%, compared to CMOS, due to the row-parallel execution of most operations. For the division operation, due to the sequential nature of \texttt{flip-flops}, CIM-based CORDIV exhibits higher latency, but it is offset by increased throughput enabled by SIMD parallelism. CORDIV is still compatible with the ADC-based S-to-B conversion, where the SBS is written as resistance values in a column, enabling the ADC to sense the bitline current that represents the number of `1's in the SBS. Other ReRAM-based implementations use the SBS as voltage input, thus requiring a reference column.

\begin{table}[b]
\centering
\caption{Hardware Cost Evaluation for \textit{CMOS-based} (\ding{59}) and \textit{ReRAM-based} (\ding{70}) SC technologies. 
}
\vspace{-0.8em}
\label{HW_costs}
\setlength{\tabcolsep}{5.5pt}
\scriptsize
\begin{tabular}{|cc|c|cc|} 
\hline
\multicolumn{3}{|c|}{\textbf{ CMOS-based Design} (\ding{59})} & \multirow{2}{*}{\begin{tabular}[c]{@{}c@{}}\textbf{ Total }\\\textbf{Latency }\textsuperscript{\ding{105}}\\ \textbf{(\SI{}{\nano\s})}\end{tabular}} & \multirow{2}{*}{\begin{tabular}[c]{@{}c@{}}\textbf{ Total }\\\textbf{Energy }\\\textbf{(\SI{}{\nano\joule})}\end{tabular}} \\ 
\cline{1-3}
\textbf{ Binary$\rightarrow$SC }\ding{182} & \multicolumn{1}{c}{\begin{tabular}[c]{@{}c@{}} \textbf{SC arithmetic}\\\textbf{operations} \ding{183}\end{tabular}} & \textbf{ SC$\rightarrow$Binary }\ding{184} &  &  \\ 
\hline
\begin{tabular}[c]{@{}c@{}}LFSR\\+\\Comparator\end{tabular} & \begin{tabular}[c]{@{}c@{}}Multiplication\\Addition\\Subtraction\\Division\end{tabular} & \multirow{2}{*}{\begin{tabular}[c]{@{}c@{}}\\ \\$log_2N$-bit \\counter\end{tabular}} & \begin{tabular}[c]{@{}c@{}}122.88 \\130.56 \\133.12 \\133.12\end{tabular} & \begin{tabular}[c]{@{}c@{}}0.23 \\0.26 \\0.16 \\0.18\end{tabular} \\ 
\cline{1-2}\cline{4-5}
\begin{tabular}[c]{@{}c@{}}Sobol\\+\\Comparator\end{tabular} & \begin{tabular}[c]{@{}c@{}}Multiplication \\Addition \\Subtraction \\Division\end{tabular} &  & \begin{tabular}[c]{@{}c@{}}125.44 \\130.56 \\133.12 \\130.56\end{tabular} & \begin{tabular}[c]{@{}c@{}}0.30 \\0.30 \\0.12 \\0.14\end{tabular} \\ 
\hline
\multicolumn{3}{|c|}{\textbf{ReRAM-based Design} (\ding{70})} & \multicolumn{2}{c|}{} \\ 
\cline{1-3}
\begin{tabular}[c]{@{}c@{}}\textit{IMSNG-opt}\end{tabular} & \begin{tabular}[c]{@{}c@{}}Multiplication \\ Addition \\Subtraction \\Division\end{tabular} &  \multirow{2}{*}{\begin{tabular}[c]{@{}c@{}} 8-bit ADC~\cite{shafiee2016isaac}\end{tabular}}  & \begin{tabular}[c]{@{}c@{}}80.8\\ 80.8\\ 81.6\\ 12544.0\end{tabular} & \begin{tabular}[c]{@{}c@{}}3.50\\3.50\\ 3.51\\ 4.48\end{tabular} \\ 
\hline
\end{tabular}
\vspace{-0.3em}
 \justify{\scriptsize {\ding{105}: Total latency=Critical Path Latency \texttimes~$N$. Bit-stream Length ($N$) is 256.}}
\end{table}

If considered in isolation, our ReRAM-based design consumes more energy than CMOS mainly due to multiple read operations ($5N$) for SBS generation and storing SBSs (at least one write operation). In this section, we only report results for the \textit{IMSNG-opt} configuration. For comparison, \textit{IMSNG-naive} requires \SI{395.4}{\nano\s} and consumes \SI{10.23}{\nano\joule} per conversion, whereas \textit{IMSNG-opt} completes the same process in \SI{78.2}{\nano\s} while using only \SI{3.42}{\nano\joule}.

\begin{figure}[t]
    \centering
        \includegraphics[width=0.9\linewidth]{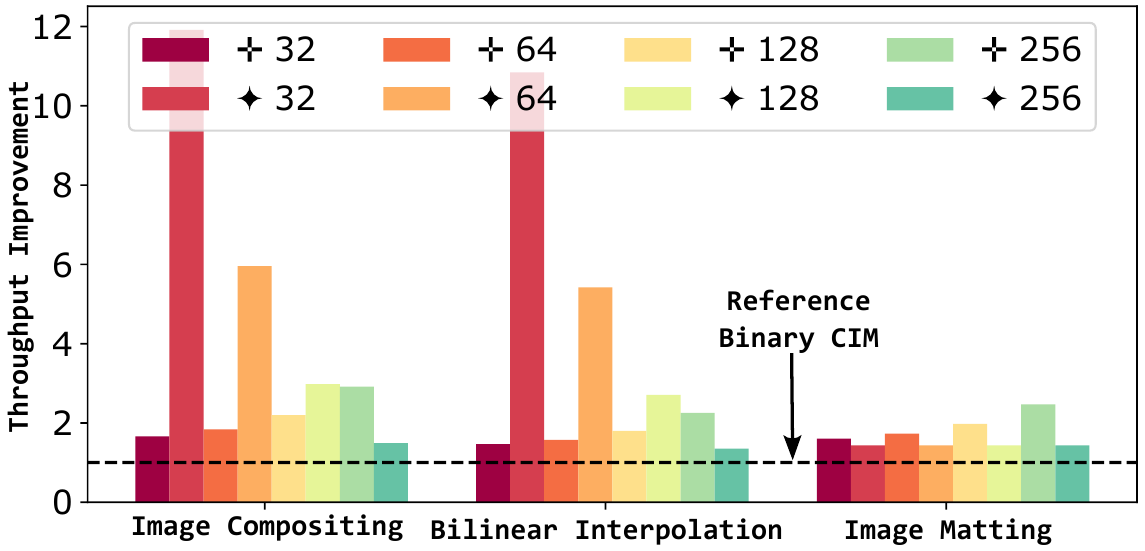}
        
    \caption{Normalized throughput for CMOS (\ding{59}) and ReRAM (\ding{70}) designs.}
    \label{fig:latency}
\end{figure}

Fig.~\ref{fig:energy} and Fig.~\ref{fig:latency} compare the SC designs (\ding{59} and \ding{70}) to the binary CIM (\ding{71}) (also considering memory transfers). 
Regarding energy savings, on average, our design reduces energy by 2.8$\times$ and 1.15$\times$, compared to the binary CIM and and CMOS designs, respectively.

Only for larger resolutions ($N$=256), our solution performs poorly compared to CMOS, but these resolutions also considerably increases latency of CMOS-based designs and are typically avoided. 
In terms of throughput, our design achieves, on average, 2.16$\times$ and 1.39$\times$ higher throughput compared to the binary CIM and CMOS designs respectively.

The off-chip communication in CMOS-based designs -- specifically loading images and storing outputs to the same ReRAM setup -- significantly increases total energy consumption. ReRAM-based designs outperform CMOS-based ones for smaller SBSs (32 and 64) across all applications and for all SBS lengths in Bilinear Interpolation. However, for larger SBSs, the cost of writing SBSs outweighs CIM's benefits to the extent that transferring data to the SC logic is more efficient.
Nonetheless, implementing SC on general-purpose CIM hardware is a key advantage, as it requires no additional components beyond those common in other CIM designs.

\begin{table}[t]
\centering 
\caption{SSIM (\%)/PSNR (dB) comparison free of (\ding{55}) or under (\ding{51}) CIM faults across different bit-stream lengths.}
\vspace{-0.8em}
\scriptsize
\setlength{\tabcolsep}{5pt}
\begin{tabular}{lcccccc}
\toprule
\multirow{2}{*}{\begin{tabular}[c]{@{}c@{}} Design \end{tabular}}  & \multicolumn{2}{c}{Image Compositing} & \multicolumn{2}{c}{Bilinear Interpolation} & \multicolumn{2}{c}{Image Matting} \\
& \ding{55} & \ding{51} & \ding{55} & \ding{51}  & \ding{55} & \ding{51}  \\
\midrule
\ding{71}~\cite{leitersdorf2023aritpim}                       &    99.9/91.8       &  82.9/42.4        &   95.6/39.0          &    64.4/37.6         &  99.9/50.3      & 4.8/-18.2       \\
\ding{70} 32                                       &    99.9/23.4      &   99.9/22.2         &   82.0/28.5          &    79.4/28.6          &  95.3/31.5      & 88.2/30.0            \\
\ding{70} 64                                        &    99.9/26.7      &   99.9/25.6          &  87.7/29.5          &    86.5/29.7        &  98.7/37.8      & 93.0/36.7              \\
\ding{70} 128                                        &    99.9/28.2      &   99.9/27.6          &  91.4/30.2          &    90.0/29.7          &  99.4/41.7      & 94.6/38.5              \\
\ding{70} 256                                        &    99.9/32.3      &   99.9/30.9          &  93.0/31.1          &      92.9/31.5        &  99.7/44.9      & 96.7/44.5              \\
\bottomrule
\label{tab:image_operations}
\end{tabular}
\justify{\scriptsize {PSNR: Peak Signal-to-Noise Ratio, SSIM: Structural Similarity. $\uparrow$ is better. For the ReRAM-based SC design(\ding{70}), different bit-stream lengths ($N$=32,64,128,256) are used.}}
\end{table}

\subsection{Reliability through Stochastic Computing}
\label{subsec:reliability}
In digital CIM, a fault is a bit flip, where results invert from the expected value.
Table~\ref{tab:image_operations} presents the quality of our selected applications \textit{with} CIM faults (realistic scenario, \ding{51}) and \textit{without} CIM faults (ideal scenario, \ding{55}).
For image compositing and bilinear interpolation, we compare the outputs of \textit{ReRAM-based SC} (\ding{70}) and \textit{Binary CIM} (\ding{71})~\cite{leitersdorf2023aritpim} against the SW implementation. For image matting, we compare the blended images obtained using the original $\alpha$ ($I$ in Fig.~\ref{fig:ImageProc}(c)) and the estimated $\hat{\alpha}$. 
Our design shows an average quality drop of 5\% under realistic scenarios.
Traditional arithmetic~\cite{leitersdorf2023aritpim} exhibits 47\% drop in quality in the presence of faults (as high as 95.2\%, in Image Matting), as errors at higher bit positions can lead to more severe and widespread inaccuracies. 
Among these, image matting relying on integer division is particularly vulnerable to faults, often rendering unacceptable outputs.

\begin{figure}[t] 
    \centering
        \includegraphics[width=0.9\linewidth]{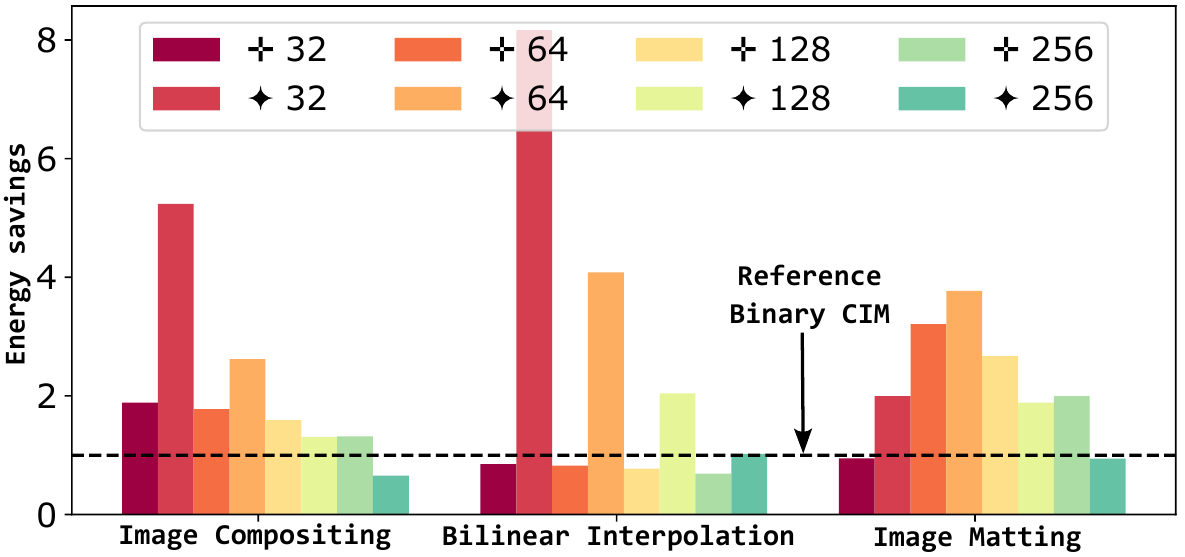}
        
        \vspace{-0.7em}
    \caption{Normalized energy savings for CMOS (\ding{59}) and ReRAM (\ding{70}) designs.}
    \label{fig:energy}
\end{figure}

Comparing different SBS resolutions shows that some algorithms require only smaller bit-streams (no noticeable drop in accuracy).
This aligns well with the performance and energy-efficiency trends presented in Fig.~\ref{fig:energy}, where energy savings increase as bit-stream size decreases.
The impact of faults is highly algorithm-dependent. For instance, image compositing and bilinear interpolation, both relying on the \texttt{MAJ} operation, are more tolerant than CORDIV to CIM faults. 
This highlights the strength of SC on unreliable CIM, offering more robustness to CIM faults and not requiring dedicated fault protection hardware, making it both energy-efficient and accurate.

Memory protection schemes exist but are costly and traditional error correction codes cannot protect CIM operations (\texttt{AND}, \texttt{OR}, and \texttt{MAJ}).
A recent work~\cite{cilasun2024error} proposes a parity-based fault-tolerance scheme for NVM-CIM, which adds a significant area overhead for parity storage and syndrome generation, as well as added latency due to critical path dependencies.
SC is fault-tolerant, suitable for unreliable devices, and does not add extra overhead to protect CIM operations.

\section{Conclusion}
\label{sec:conclusions}
We presented a ReRAM-based, in-memory implementation of stochastic computing (SC). Leveraging existing in-ReRAM true RNGs, we produce SBSs, perform stochastic operations, and convert them back to binary -- all within the memory array. Compared to the state-of-the-art CMOS-based SC solution, our results across multiple image processing kernels show similar accuracy, $1.39\times$ higher throughput and $1.15\times$ less energy consumption, all without requiring specialized logic for SC. 

\section*{Acknowledgments}
This work is partially funded by National Science Foundation (NSF) under grant No. 2019511, 2339701, the German Research Council (DFG) through the HetCIM project (502388442), the AI competence center ScaDS.AI Dresden/Leipzig (01IS18026A-D), the CRC/TRR 404-Active 3D, and generous gifts from NVIDIA. We thank Stefan Wiefels from Forschungszentrum Jülich for our insightful discussions on the earlier version of this paper. 

\bibliographystyle{IEEEtran}
\renewcommand{\bibfont}{\footnotesize}
\bibliography{sc}
\end{document}